\documentclass{jfm}

\usepackage{graphicx}
\usepackage{newtxtext}
\usepackage{newtxmath}
\usepackage{natbib}
\usepackage{hyperref}
\hypersetup{
    colorlinks = true,
    urlcolor   = blue,
    citecolor  = black,
}

\newcommand{\RomanNumeralCaps}[1]
\renewcommand{\Pr}{\Pran}
\newcommand{\Ra}{\mbox{\textit{Ra}}}
\newcommand{\Ro}{\mbox{\textit{Ro}}}
\newcommand{\iRo}{\mbox{\textit{Ro}}^{-1}}

\newcommand{\Nu}{\mbox{\textit{Nu}}}

\title{Interplay between advective, diffusive, and active barriers in Rayleigh-B\'enard flow}

\author{Nikolas O. Aksamit\aff{1} \corresp{\email{nikolas.aksamit@uit.no}},
 Robert Hartmann\aff{2},
Detlef Lohse \aff{2,3},
George Haller \aff{4}}

\affiliation{\aff{1} Institute for Mathematics and Statistics, UiT - The Arctic University of Norway, 9037 Troms\o, Norway
\aff{2} Physics of Fluids Group and Max Planck Center for Complex Fluid Dynamics, J. M. Burgers Centre for Fluid Dynamics, University of Twente, 7500AE Enschede, The Netherlands
\aff{3} Max Planck Institute for Dynamics and Self-Organization, 37077 Göttingen, Germany
\aff{4} Institute for Mechanical Systems, ETH Zürich, 8092 Zürich, Switzerland} 

\begin{document}
\maketitle

\begin{abstract}
Our understanding of the material organization of complex fluid flows has recently benefited from mathematical developments in the theory of objective coherent structures. These methods have provided a wealth of approaches that identify transport barriers in three-dimensional (3D) turbulent flows. Specifically, theoretical advances have been incorporated into numerical algorithms that extract the most influential advective, diffusive and active barriers to transport from data sets in a frame-indifferent fashion. To date, however, there has been very limited investigation into these objectively-defined transport barriers in 3D unsteady flows with complicated spatiotemporal dynamics. Similarly, no systematic comparison of advective, diffusive and active barriers has been carried out in a 3D flow with both thermal and shear-driven features. In our study, we utilize simulations of turbulent rotating Rayleigh-Bénard convection to uncover the interplay between advective transport barriers (Lagrangian coherent structures), material barriers to diffusive heat transport and objective Eulerian barriers to momentum transport. For a range of (inverse) Rossby numbers, we visualize each type of barrier and quantify their role as barriers to momentum and heat transport under changes in the relative influence of mechanical and thermal forces.

\end{abstract}

{\bf MSC Codes }  {\it(Optional)} Please enter your MSC Codes here

\section{Introduction}
\label{sec:intro}

Global transport properties in turbulent flows are intimately connected to flow structure and flow organization. 
Clearly, this holds for thermally driven turbulent flows, where the key global transport property is the 
heat transport. This is of utmost importance in many technological applications and in the natural flows one finds in the ocean, in the atmosphere, and in the interior of stars and planets. The paradigmatic system for thermally driven flow is Rayleigh-B\'enard 
(RB) convection
\citep{bod00,kad01,ahl09,loh10,chi12,shishkina2021}, a fluid in a box heated from below and cooled from above.
The key question for this flow is: How does the heat transfer (the Nusselt number) depend on the control parameters, such as the non-dimensional temperature difference between
top and bottom plates (the Rayleigh number), the ratio of kinematic viscosity and thermal diffusivity (the Prandtl number), and the 
the ratio between width(s) of the box and its height? The same fundamental question holds for rotating RB convection when one varies the rate of rotation \citep{ecke2023}, as expressed by the inverse Rossby number.
 
 Turbulent RB flow is characterized by the interplay between large scale convection
 rolls, thermal boundary layers, and plumes detaching from these boundary layers. The interplay between these
 flow structures determines the overall heat flux. Any model for the heat transport in RB flow makes assumptions
 on the flow structures. The unifying theory of thermal convection of Grossmann and Lohse \citep{gro00,gro01,ahl09,ste13} successfully describes the functional dependence of the Nusselt number ($Nu$) and the Reynolds number on the Rayleigh number and the Prandtl number. Such connections must also be made in the theories for rotating RB flow, where the rotation 
 first leads to an increase in heat transport and then a strong decrease, due to the Taylor-Proudman theorem (see the review article by \citet{ecke2023}). The increase in $Nu$ as function of rotation rate for intermediate Rossby numbers can be understood from 
 the overall flow organization and its connection to the thermal boundary layers \citep{zho09t,ste09}, a concept which
 can be extended to laterally confined RB convection \citep{cho15,cho16,hartmann2021} and to double diffusive convection \cite{cho17}. 
 \citet{hartmann2022} later extended this concept to account for the heat transport in laterally confined and rotating RB flow, connecting the overall heat flux with both the boundary layer structures and the flow organization in the bulk.

 The flow organization and the flow structure are not only relevant for the time-averaged heat flux, but also
 for its temporal evolution. \citet{nik05}, \citet{bro06}, \citet{xi07}, \citet{zwirner2020}, and \citet{shishkina2021}
 showed that rearrangements of the roll structure over time lead to correlated modifications in the heat flux. 
 Each roll is quite isolated from its neighbors and has a long-lasting identity. Similarly, \citet{sug10} showed that the reversal of the large scale convection roll is related to the spatial growth of otherwise spatially isolated corner flows. The reversal is then correlated with a burst in the temporal evolution of $Nu$ as the corner flow grows to the size of the system and takes over the role of the large scale roll. In both cases, we use the term isolated to refer to minimal exchange of momentum and heat with the surrounding flow, as if there exists a partial flow barrier there. 
 
 Given the important role of flow organization for the overall heat transport, quantitative criteria are needed
 to characterize the flow structure. In their 2D numerical simulations, \citet{sug10} used the overall angular velocity
 to achieve this. In 3D RB convection, the $Q$-criterion and $\lambda_2$ criterion \citep{Hunt1988,Jeong1995} have been previously used \citep{bou86,vor98,vor02,kun10,wei10}. Though these methods have their merits and are Galilean invariant, they are {\it not objective}, i.e. give different results in general moving frames. Results returned by these criteria, therefore, depend on the observer and generally do not capture properties and structures intrinsic to the flow. Furthermore, the original, physically motivated values of these criteria are generally ignored and user-selected level surfaces of the quantities involved are presented instead. As a result, a precise definition for coherent structures is missing in the applications of these methods.
 
 Fortunately, over the last two decades, mathematics methods for the objective identification and characterization of material fluid structures and transport barriers have been developed \citep{hal05,Haller2015,Haller2016,Haller2020,Haller2023}. These methods uncover experimentally verifiable flow features that are indifferent to the choice of the frame of reference. Technically, this means that the results of 
these structure or barrier
identification procedures do not change under time-varying rotation and translation frame transformations of the form
$$\mathbf{x}=\mathbf{Q}(t)\mathbf{y}+\mathbf{b}(t),$$
where
$\mathbf{Q}(t)$ is an arbitrary, time-dependent rotation matrix and
$\mathbf{b}(t)$ is an arbitrary time-dependent translation vector.
This objectivity requirement ensures that  transport barriers 
 are intrinsic to the flow and can be unambiguously visualized
experimentally using material tracers or dye. As shown, the $Q$-criterion and $\lambda_2$ criterion do not pass this objectivity test \citep{Haller2015}.

 \begin{figure}
\centerline{\includegraphics[width=5in]{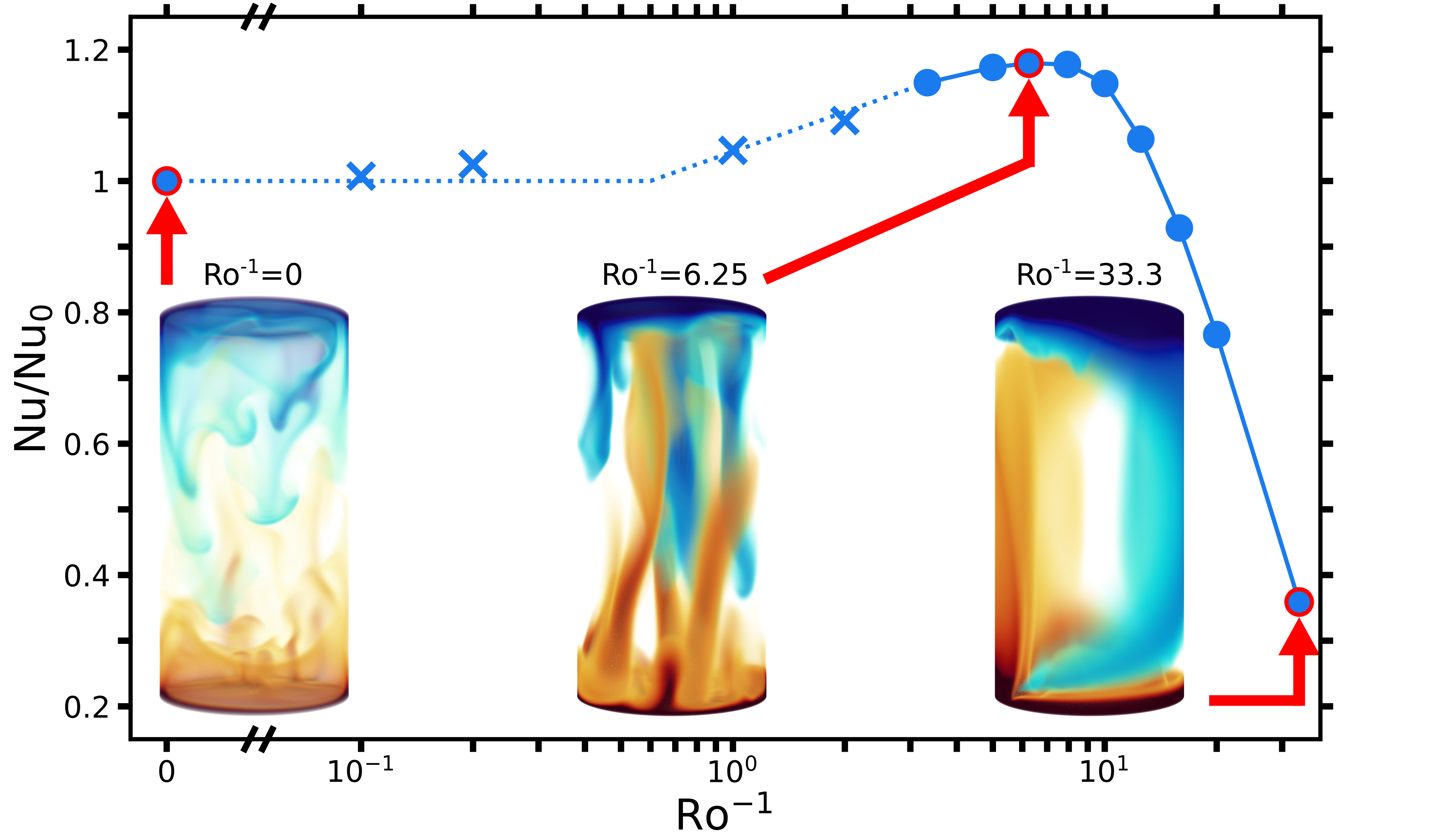}}% Images in 100% size
\caption{Normalized heat transport $\Nu/\Nu_0$ with respect to the inverse Rossby number $\iRo$ for $\Pr=4.38$ and $\Ra=2\times10^8$ in a $\Gamma=0.5$ cylinder. The circles show data points published in \citet{hartmann2022}, the crosses belong to unpublished data. The dotted line schematically shows the expected behavior for the onset of heat transport enhancement. The arrows indicate the three cases considered in this paper. The volume renders of the instantaneous temperature field illustrate the typical, very different flow morphology for these cases.}
\label{fig:NuVsRo}
\end{figure}
 
%\subsection{Objectively defined passive and active transport barriers in 3D flows}

The purpose of the present research is to investigate objectively-defined heat and momentum barriers as well as rotationally coherent  Lagrangian coherent structures in non-rotating and rotating RB flow. This approach allows us to better identify the flow organization, the flow structures, vortices, and transport barriers and how they develop
under varying control parameters (Rayleigh and Rossby numbers). To achieve this, we utilize direct numerical simulations of RB 
flow in a cylinder of aspect ratio $\Gamma = 1/2$ at a Rayleigh number of $Ra = 2\times 
10^8$ 
and three different (inverse) Rossby numbers, 
namely $Ro^{-1} = 0$ (no rotation), 
$Ro^{-1} = 6.25$ (intermediate rotation, close to the maximum in heat transfer), and 
$Ro^{-1} = 33.3$ (strong rotation, with strong suppression of the heat transfer
due to the Taylor-Proudman theorem). A comparison of these three scenarios is shown in figure \ref{fig:NuVsRo}.

The paper is organized as follows: To be self-contained, we briefly discuss passive, diffusive, and active transport barriers in 
Section \ref{barriers}. For a detailed and more general discussion we refer the reader to \cite{Haller2023}. 
In Section \ref{dyn-eqs}, we give the underlying equations of rotating RB convection and the numerical method we 
have employed for the direct numerical simulations of these equations. 
Section 
\ref{results}
presents the results, which are discussed in section \ref{discuss}. The paper ends with conclusions 
and an outlook (Section \ref{conclu}).

\section{Different types of transport barriers}\label{barriers}

Transport barriers inhibit the spread of material or specific quantities
associated with a fluid flow. We will refer to the inhibitors of
the spread of material (or of a conserved tracer field) as \emph{advection
barriers}, whereas we will call the inhibitors of the transport of
diffusive tracer fields \emph{diffusion barriers}. Both advection
and diffusion barriers are assumed to be passive here, which means
that their evolution does not change the fluid velocity field. In
contrast, we refer to the transport of quantities with a direct connection
to the fluid velocity field as active transport, with examples including
the transport of vorticity and linear momentum. Barriers inhibiting
the spread of an active field will be referred to as \emph{active
barriers}. For a general discussion of all these barrier types and
for ways to detect them, we refer again to \cite{Haller2023}.

\subsection{Advection barriers}

Defining the observed barriers to material advection precisely is challenging
as all material surfaces block the transport of conserved tracers across them. The
concept of Lagrangian Coherent Structures (LCS) addresses this ambiguity
by seeking distinguished material surfaces that are the centerpieces
of material deformation and maintain coherence over some finite time interval of interest
(\cite{Haller2015,Haller2023}). \emph{Hyperbolic LCSs} are defined as local maximizers of
repulsion or attraction among material surfaces in the flow. In contrast,
\emph{elliptic LCSs} are defined as local maximizers of shear among
material surfaces. 

In a given velocity data set $\mathbf{v}(\mathbf{x},t)$, LCS detection tools use fluid particle trajectories $\mathbf{x}(t;t_{0},\mathbf{x}_{0})$ generated from the differential equation
\begin{equation}
\dot{\mathbf{x}}=\mathbf{v}(\mathbf{x},t)\label{eq:fluid flow},
\end{equation} 
with initial position $\mathbf{x}_{0}$ at time $t_{0}$. These trajectories define the flow map \textbf{$\mathbf{F}_{t_{0},t}\left(\mathbf{x}_{0}\right)=\mathbf{x}(t;t_{0},\mathbf{x}_{0})$}, from which we also define the right Cauchy--Green strain tensor
\[
\mathbf{C}_{t_{0},t}=\left[\boldsymbol{\nabla}\mathbf{F}_{t_{0},t}\right]^{\mathrm{T}}\boldsymbol{\nabla}\mathbf{F}_{t_{0},t}.
\]
 
To visualize hyperbolic LCSs, it is common to use the \emph{finite-time Lyapunov
exponent} (FTLE) field, defined over a finite time interval $\left[t_{0},t_{1}\right]$
as
\begin{equation}
\mathrm{FTLE}_{t_{0},t_{1}}(\mathbf{x}_{0})=\frac{1}{2\left|t_{1}-t_{0}\right|}\log\lambda_{\mathrm{max}}\left(\mathbf{C}_{t_{0},t_{1}}\left(\mathbf{x}_{0}\right)\right),\label{eq:FTLE field}
\end{equation}
where $\lambda_{\mathrm{max}}>0$ is the largest eigenvalue of the positive definite tensor $\mathbf{C}_{t_{0},t_{1}}$.

FTLE values measure locally the largest material stretching rate in
the flow. For $t_{1}-t_{0}$ large enough, codimension-one $\mathrm{FTLE}_{t_{0},t_{1}}(\mathbf{x}_{0})$ ridges align with $t_{0}$
positions of maximally repelling LCSs. Similarly, for $t_{0}-t_{1}$ large enough (backward time integration), we can find maximally attracting LCSs surfaces \citep{Haller2015,Haller2023}).

To detect elliptic LCSs, we define the \emph{Lagrangian-averaged
vorticity deviation} (LAVD, \citep{Haller2016}), by measuring the average deviation of the
pointwise vorticity $\boldsymbol{\omega}=\boldsymbol{\nabla}\times\mathbf{v}$
from its spatial mean $\bar{\boldsymbol{\omega}}$. Specifically, we have the objective Lagrangian
diagnostic 
\begin{equation}
\mathrm{LAVD}_{t_{0},t_{1}}(\mathbf{x}_{0}):=\frac{1}{|t_{1}-t_{0}|}\int_{t_{0}}^{t_{1}}\left|\boldsymbol{\omega}(\mathbf{F}_{t_{0},s}(\mathbf{x}_{0}),s)-\bar{\boldsymbol{\omega}}(s)\right|ds.\label{eq:LAVD}
\end{equation}
This quantity is objective once
the fluid mass involved in the spatial averaging is fixed but the
result will depend on the choice of that domain. In the Rayleigh-B\'enard
setting we analyze here, the domain will be simply fixed as the full
computational domain. 

Locations of \emph{elliptic LCSs} at time $t_{0}$ can be
identified as smooth cylindrical level surfaces of \emph{$\mathrm{LAVD}_{t_{0},t_{1}}(\mathbf{x}_{0})$}
surrounding a unique, codimension-two ridge \citep[see][]{Neamtu-Halic2019,Haller2023}. In turn\emph{, coherent Lagrangian vortices} can be located as nested
families of such elliptic LCSs. 

\subsection{Diffusion barriers}

In contrast to advective barriers, diffusive barriers can be defined
unambiguously without any reliance on a coherence definition. If a scalar $c$ satisfies the classic advection-diffusion equation with with
diffusivity $\kappa>0$, then the diffusive transport through an evolving
material surface $\mathcal{M}(t)$ with $\mathcal{M}(t_{0})=\mathcal{M}_{0}$
can be written
\begin{equation}
\Sigma_{t_{0},t_{1}}\left(\mathcal{M}_{0}\right)=\int_{t_{0}}^{t_{1}}\int_{\mathcal{M}(t)}\kappa\mathbf{\boldsymbol{\nabla}}c\cdot\mathbf{n}\,dA\,dt,\label{eq:transport1}
\end{equation}
with $\mathbf{n}(\mathbf{x},t)$ denoting a smoothly oriented unit
normal vector field along $\mathcal{M}(t)$. \cite{Haller2018} sought to minimize the functional $\Sigma_{t_{0},t_{1}}\left(\mathcal{M}_{0}\right)$ and found that
diffusive transport minimizers are marked by ridges
of the \emph{diffusion barrier sensitivity} (DBS\emph{)} field, defined
as 
\begin{equation}
\mathrm{DBS}_{t_{0},t_{1}}(\mathbf{x}_{0})\coloneqq\mathrm{Trace}\left[\frac{1}{|t_{1}-t_{0}|}\int_{t_{0}}^{t_{1}}\mathbf{C}_{t_{0},t}(\mathbf{x}_{0})\,dt\right].\label{eq:DBS def}
\end{equation}
Similarly, the time $t_{0}$ positions of diffusion
maximizers should be close to ridges of $\mathrm{DBS}_{t_{1}}^{t_{0}}(\mathbf{x}_{0})$.
Remarkably, DBS a predictive field given that its computation requires no diffusive simulation and only relies on the velocity field. 

For the advective and diffusive barriers evaluated in this study, we use the inverse flow map, beginning at $t_0$ and integrating backward in time integration. When plotting LAVD at $t_0$ in this way, we reveal the rotational behavior of the fluid immediately prior to $t_0$, and identify the advective transport barriers that are organizing the scalar fields at $t_0$. DBS ridges computed using this backward time integration are barriers that maximize diffusive heat transport and are detail Lagrangian structures that maintain strong temperature gradients through their boundaries.

\subsection{Momentum barriers}

Defining transport barriers objectively for active dynamical quantities
is a challenge because the most often used such quantities (such as
the vorticity and the momentum) are not objective. To circumvent this
problem, \cite{Haller2020} introduce the (average) \emph{diffusive
transport} of the linear momentum vector $\mathbf{f}\left(\mathbf{x},t\right)=\rho\mathbf{v}(\mathbf{x},t)$
through $\mathcal{M}(t)$, defined over a time interval $\left[t_{0},t_{1}\right]$
as
\begin{equation}
\psi_{t_{0},t_{1}}\left(\mathcal{M}_{0}\right)=\frac{1}{t_{1}-t_{0}}\int_{t_{0}}^{t_{1}}\left[\int_{\mathcal{M}(t)}\frac{D\mathbf{f}}{Dt}\cdot\mathbf{n}\,dA\right]_{vis}dt=\frac{\nu\rho}{t_{1}-t_{0}}\int_{t_{0}}^{t_{1}}\int_{\mathcal{M}(t)}\Delta\mathbf{v}\cdot\mathbf{n}\,dAdt,\label{eq:internal flux}
\end{equation}
where the $\left[\,\cdot\,\right]_{vis}$ operation identifies the
part of the bracketed quantity that has an explicit dependence on
the viscosity $\nu$, as determined from the incompressible Navier--Stokes
equation with constant density $\rho$. Barriers to
momentum transport (or momentum barriers for short) can be defined
as material surfaces whose initial positions $\mathcal{M}_{0}$ are
local extremizers of the functional $\psi_{t_{0},t_{1}}\left(\mathcal{M}_{0}\right)$.
Of these extremizers, the strongest ones are \emph{active momentum
barriers}, which are structurally stable surfaces admitting strictly
zero transport over any of their subsets. 

In the instantaneous limit, $t_{1}\to t_{0}=t$, a surface $\mathcal{M}(t)$ is 
a perfect barrier to momentum flux if $\Delta\mathbf{v}\cdot\mathbf{n}$ vanishes at each point of $\mathcal{M}(t)$.  Finding \emph{active Eulerian (instantaneous) momentum barriers} then reduces to finding structurally stable invariant manifolds in the barrier vector field 

\begin{equation}
\mathbf{x}^{\prime}=\Delta\mathbf{v}(\mathbf{x},t_0) \label{eq:active barrier}
\end{equation}
at a given time $t_0$. Here prime denotes differentiation with respect to the barrier time $s$ that parameterizes trajectories of (\ref{eq:active barrier}) that form barrier surfaces.
The barrier vector field is a steady,
volume-preserving dynamical system because $\Delta\mathbf{v}$ is
divergence-free for incompressible flows. For this vector field, we can compute FTLE and LAVD fields from trajectories defined by the instantaneous barrier field flow map \textbf{$\mathcal{F}_{(\mathbf{x}_{0}; t_{0})}^{s}=\mathbf{x}(s;\mathbf{x}_{0};t_{0})$} where $\mathbf{x}(s;\mathbf{x}_{0};t_{0})$ is a trajectory of (\ref{eq:active barrier}) computed for $t_0$ fixed. In all our computations, we select the integration length $s$ for $\mathcal{F}_{(\mathbf{x}_{0}; t_{0})}^{s}$ to be on the scale of the decorrelation timescale described by \citet{Aksamit2022}. Beyond this time scale, trajectory-based diagnostics become less informative about structures near their initial positions, $\mathbf{x}_{0}$. Hyperbolic and elliptic invariant manifolds of (\ref{eq:active barrier}) are detected in analogy with their advective counterparts in the \emph{active FTLE} (aFTLE) fields,
\begin{equation}
\mathrm{aFTLE}_{t_{0}}^{s}(\mathbf{x}_{0})=\frac{1}{2\left|s\right|}\log\lambda_{\mathrm{max}}\left(\boldsymbol{\mathcal{C}}_{t_{0}}^{s}\left(\mathbf{x}_{0}\right)\right).\label{eq:aFTLE -- Lagrangian}
\end{equation}
and the \emph{active LAVD} (aLAVD) fields
\begin{equation}
\mathrm{aLAVD}_{t_{0}}^{s}(\mathbf{x}_{0}):=\frac{1}{s}\int_{0}^{s}\left|\mathbf{w}(\boldsymbol{\mathcal{F}}_{(t_{0},\mathbf{x}_{0})}^{\tilde{s}}\left(\mathbf{x}_{0}\right))-\bar{\mathbf{w}}\right|d\tilde{s}.\label{eq:active LAVD}
\end{equation}
Here $\boldsymbol{\mathcal{C}}$ is the active Cauchy-Green strain tensor generated by the active flow map $\mathcal{F}$ and $\mathbf{w}$ is vorticity of the active barrier field $\Delta\mathbf{v}$.

\section{Dynamical equations and numerical method}\label{dyn-eqs}

Rotating RB convection is governed by a set of equations including the continuity equation, the Navier-Stokes equations and the convection-diffusion equation of temperature. Under the Oberbeck-Boussinesq approximation they are given in their dimensionless form as
\begin{align}
\label{eq:RRBC1}
\nabla\cdot\mathbf{v} &= 0, \\
\frac{\mathrm{d}\mathbf{v}}{\mathrm{d}t} &= -\nabla P + \sqrt{\frac{\Pr}{\Ra}}\Delta\mathbf{v} + \Theta \vec{e}_z - \frac{1}{\Ro}\vec{e}_z\times\mathbf{v},\\
\frac{\mathrm{d}\Theta}{\mathrm{d}t} &= \frac{1}{\sqrt{\Pr \Ra}}\Delta\Theta.
\label{eq:RRBC3}
\end{align}
Therein, $\mathbf{v}$, $P$ and $\Theta$ are the dimensionless velocity, pressure and temperature fields, respectively, normalized by the height $H$ between the plates and the free-fall velocity $U_0=\sqrt{\alpha g \delta T H}$, where $\alpha$ is the isobaric thermal expansion coefficient, $g$ the gravitational acceleration and $\delta T$ the temperature difference between the upper and lower plates. The pressure field $P$ is further reduced by the hydrostatic balance and centrifugal contributions. Hence, the set of equations depends on three control parameters: the Prandtl number $\Pr=\nu/\kappa$, the Rayleigh number $\Ra=\alpha g \delta T H^3/(\nu\kappa)$ and the inverse Rossby number $\iRo=2\Omega H/U_0$, where $\nu$ is the kinematic viscosity, $\kappa$ the thermal diffusivity and $\Omega$ the rotation rate. Time will be measured in terms of the free fall timescale $t_{\!f\!\!f}=H/U_0$.

The RB system (\ref{eq:RRBC1})-(\ref{eq:RRBC3}) is confined to a cylinder of diameter to height ratio $\Gamma=0.5$ with no slip boundaries at the plates and the sidewall. The top and bottom plates are isothermal with $\Theta=0$ and $\Theta=1$, respectively, whereas the sidewall is adiabatic. In our simulations, we keep $\Pr=4.38$ and $\Ra=2\times10^8$ fixed, and consider the cases for no rotation ($\iRo=0$), "optimal" intermediate rotation ($\iRo=6.25$) and rapid rotation ($\iRo=33.\overline{3}$, see Fig.~\ref{fig:NuVsRo} and \citet{hartmann2022}).\\

We solve equations (\ref{eq:RRBC1})-(\ref{eq:RRBC3}) by using a central second-order accurate finite-difference scheme on a staggered grid \citep[see][]{verzicco_finite-difference_1996, verzicco_transitional_1997, verzicco_prandtl_1999}. The computational domain consists of $N_\vartheta \times N_r\times N_z=384\times64\times256$ grid points in the azimuthal, radial and vertical directions, respectively. The grid points are further clustered towards the plates and the sidewall to ensure a sufficient resolution of the boundary layers \citep{shishkina_boundary_2010} and the Kolmogorov scales in the entire domain. More details about the simulations can be found in \citet{hartmann2022}, on which the three analyzed cases of this study are grounded on. For our analysis, we consider a period of 25 free-fall times in the statistically stationary regime ($800\leq t/t_{\!f\!\!f}\leq825$). We note that the 3D flow fields are horizontally interpolated to a Cartesian grid in the post-processing to apply the different barrier diagnostics.

\section{Results}\label{results}

\begin{figure}
\centerline{\includegraphics[width=6in]{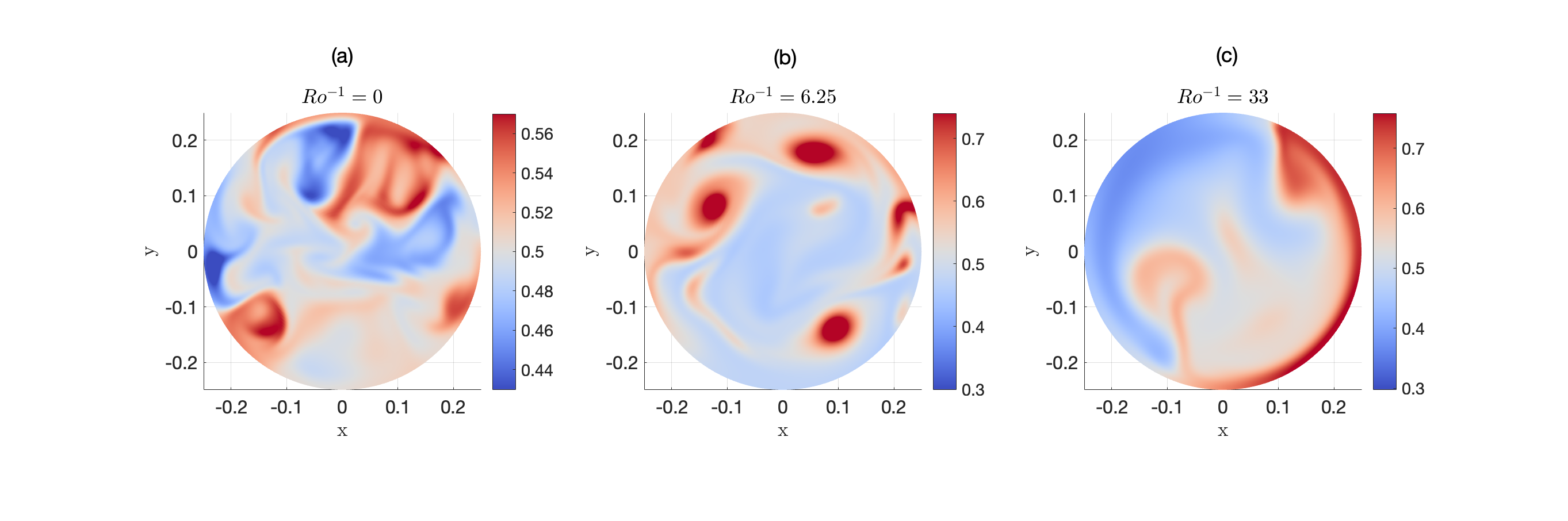}}% Images in 100% size
\caption{Comparison of instantaneous snapshots of temperature in RB convection for three different Rossby numbers. Panel (a) shows $\Theta$ at $z/H=0.5$, whereas panels (b) and (c) reveal structures closer to the heat source at $z/H=0.1$. All colormaps are centered around the plane-averaged temperature $\left<\Theta\right>_\mathcal{H}$ at that height. Classical non-rotating RB convection, Ekman pumping in vertically aligned vortices, and sidewall boundary flow characterizes the flow in panels (a)-(c), respectively.}
\label{fig:Theta_Fields}
\end{figure}

We investigated the role of transport barriers in RB convection for three different strengths of rotation, representing three distinct flow regimes. In figure \ref{fig:Theta_Fields}, we show horizontal slices of the temperature field $\Theta$ at differing heights for our three different simulations. Figure \ref{fig:Theta_Fields}a intersects our flow volume at $z/H=0.5$, whereas Figs. \ref{fig:Theta_Fields}b-c show intersections at $z/H=0.1$. We use visualizations closer to the lower boundary for our two rotational cases as turbulence in the center of the column was dampened by the Coriolis force. Further quantitative analysis will utilize the entire volume for all three flows.

The temperature fields indicate the presence of distinct flow structures and organization. Figure \ref{fig:Theta_Fields}a reveals highly turbulent mixing for $Ro^{-1}=0$. In figure \ref{fig:Theta_Fields}b, three distinct hot vortices reveal evidence of ongoing Ekman pumping at $Ro^{-1}=6.25$. At the highest rate of rotation, $Ro^{-1}=33$ in figure \ref{fig:Theta_Fields}c, the flow is much less turbulent, with the dominant features along the sidewall depicting the dominant roll of wall modes. We investigate the agreement of these signatures with advective, diffusive, and momentum barriers, as well as provide additional insights from our transport barrier focused approach, in the following sections.

\subsection{$\iRo=0$}
\label{iRo=0}

\begin{figure}
\centerline{\includegraphics[width=5in]{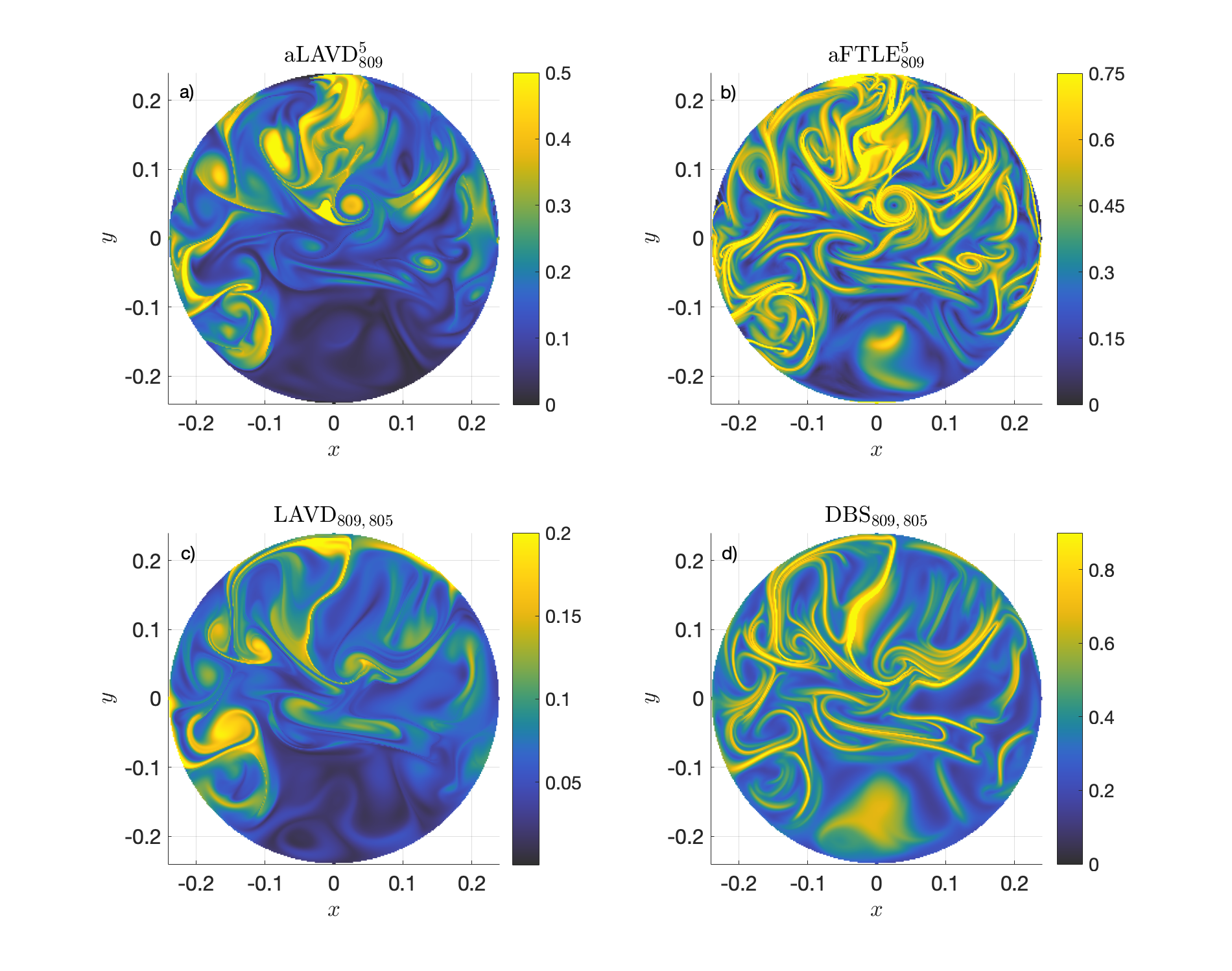}}% Images in 100% size
\caption{Comparison of advective, diffusive and active barriers for $\mathrm{Ro}^{-1} = 0$ at $t/t_{\!f\!\!f}=809$. Increased complexity of Eulerian active barriers, but very similar organization of heat, momentum, and fluid.}
\label{fig:Ro Inf}
\end{figure}

For the strongly convective case with no rotation ($Ro^{-1}=0$), our transport barrier diagnostics reveal a complex network of hyperbolic and elliptic structures both in the fluid velocity field, and in the momentum barrier field. In figure \ref{fig:Ro Inf}, we plot aLAVD, aFTLE, LAVD, and DBS fields at $z/H=0.5$ at the same dimensionless flow time ($t/t_{\!f\!\!f}=809$) as that visualized in figure \ref{fig:Theta_Fields}. 

Figures \ref{fig:Ro Inf}a and \ref{fig:Ro Inf}b reveal instantaneous momentum barriers in aLAVD and aFTLE fields calculated from an active barrier field integration time of $s=5$. We remain consistent with this integration time for all values of $\iRo$. Many of the same organizing structures can be identified in both the aLAVD and aFTLE fields. For example, rotationally coherent momentum barriers appear as both concentric families of circular aLAVD level sets around maxima and regions encircled by aFTLE ridges. 

Separations between hot and cold plumes in figure \ref{fig:Theta_Fields} clearly correlate with momentum barriers, but the momentum barriers are actually much more complex than one would expect from the temperature field alone. In the region of $y>0$, there is one such robust hot-cold interface, in Fig. \ref{fig:Theta_Fields}. Both aLAVD to aFTLE highlight structures parallel to $\Theta=\left<\Theta\right>_\mathcal{H}$, but multiple spiraling features are also revealed in adjacent regions with significantly weaker temperature gradients. The vortex which appears to mix warm and cold regions at approximately $(x,y)=(0,0.5)$ will be investigated further in Section \ref{discuss}. Further qualitative comparisons reveal many additional detailed structures in aLAVD and aFTLE, whereas a seemingly low-pass filtered version of such structure follows the temperature gradients.

Figure \ref{fig:Ro Inf}c show the intersection of coherent Lagrangian vortices with the $z/H=0.5$ plane. In contrast to the active barrier field approach used for identifying barriers in figures \ref{fig:Ro Inf}a and \ref{fig:Ro Inf}b, these rotational structures are generated by advecting fluid particles in the time-varying fluid velocity field. We use the inverse flow map, beginning at $t_0$ and integrating backward in time integration. As mentioned before, calculating LAVD this way reveals the rotational behavior of the fluid immediately prior to $t_0$, and identifies the advective transport barriers that are organizing the scalar fields at $t_0$. Figure \ref{fig:Ro Inf}d complements this analysis and shows barriers that maximize diffusive heat transport as DBS ridges from the inverse flow map.

Similarly to our comparison with momentum barriers, the advective and diffusive barriers also show a general agreement with the scalar distribution in \ref{fig:Theta_Fields}a. This confirms that DBS is indeed a predictive field as its computation relies solely on the velocity field, requiring no diffusive simulation. There is strong correlation between DBS and LAVD, and they both reveal more detail about the internal structure of advective and diffusive barriers than one sees in $\Theta$. For example, there are multiple double-plume structures, such as at $(x,y)=(-0.1,\pm 0.1)$, whose vortices are not clearly defined in $\Theta$. There is also a difference between the instantaneous momentum barrier features and the Lagrangian structures, such as the double-plume at $(x,y)=(-0.1, -0.1)$. These differences will be explored in more detail in the Section \ref{discuss}.
 
 \begin{figure}
\centerline{\includegraphics[width=5in]{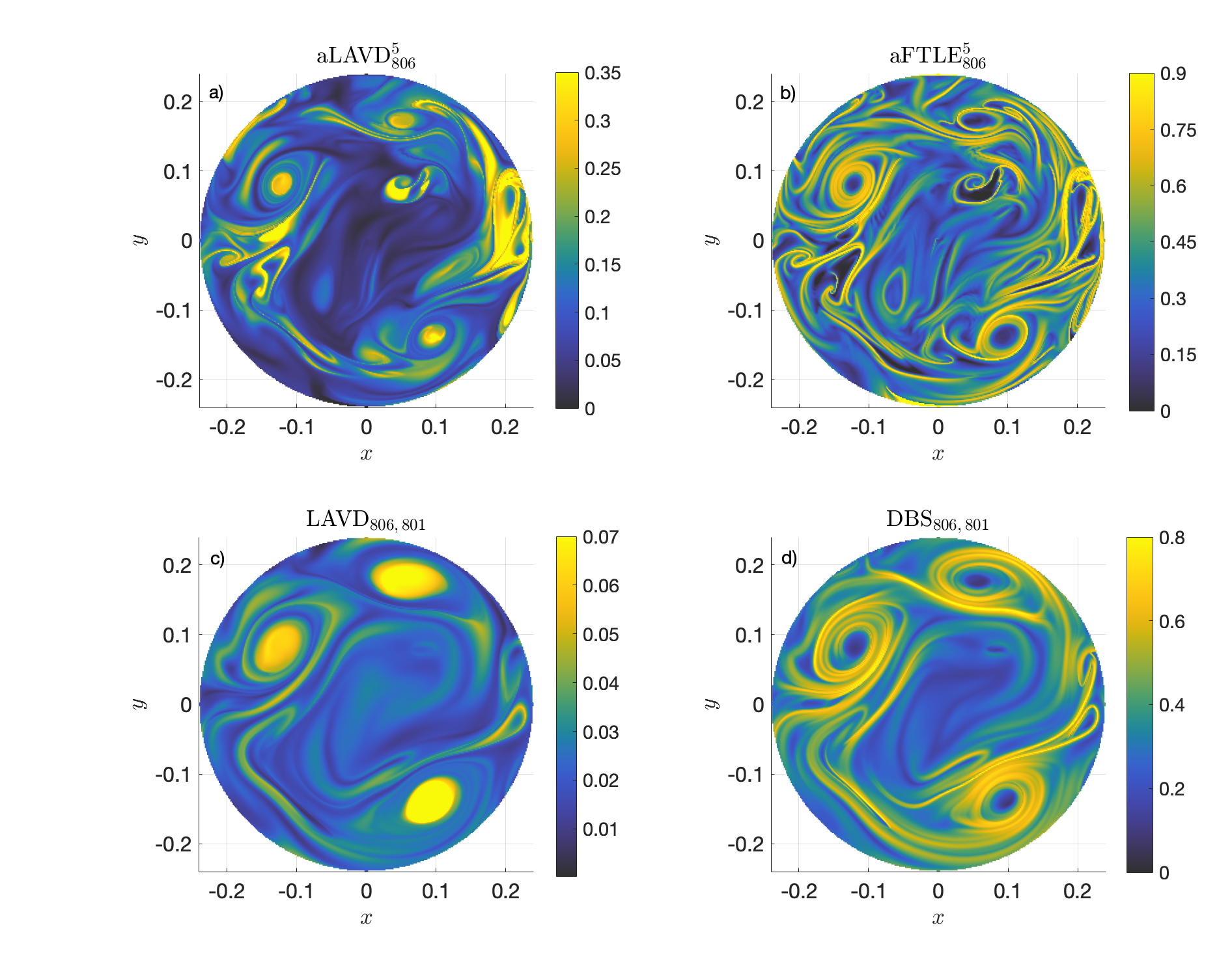}}% Images in 100% size
\caption{Comparison of advective, diffusive and active barriers for $\mathrm{Ro}^{-1} = 6.25$ at $t/t_{\!f\!\!f}=806$, emphasizing the Ekman vortices in the flow.}
\label{fig:Ro 16}
\end{figure}

 \begin{figure}
\centerline{\includegraphics[width=5in]{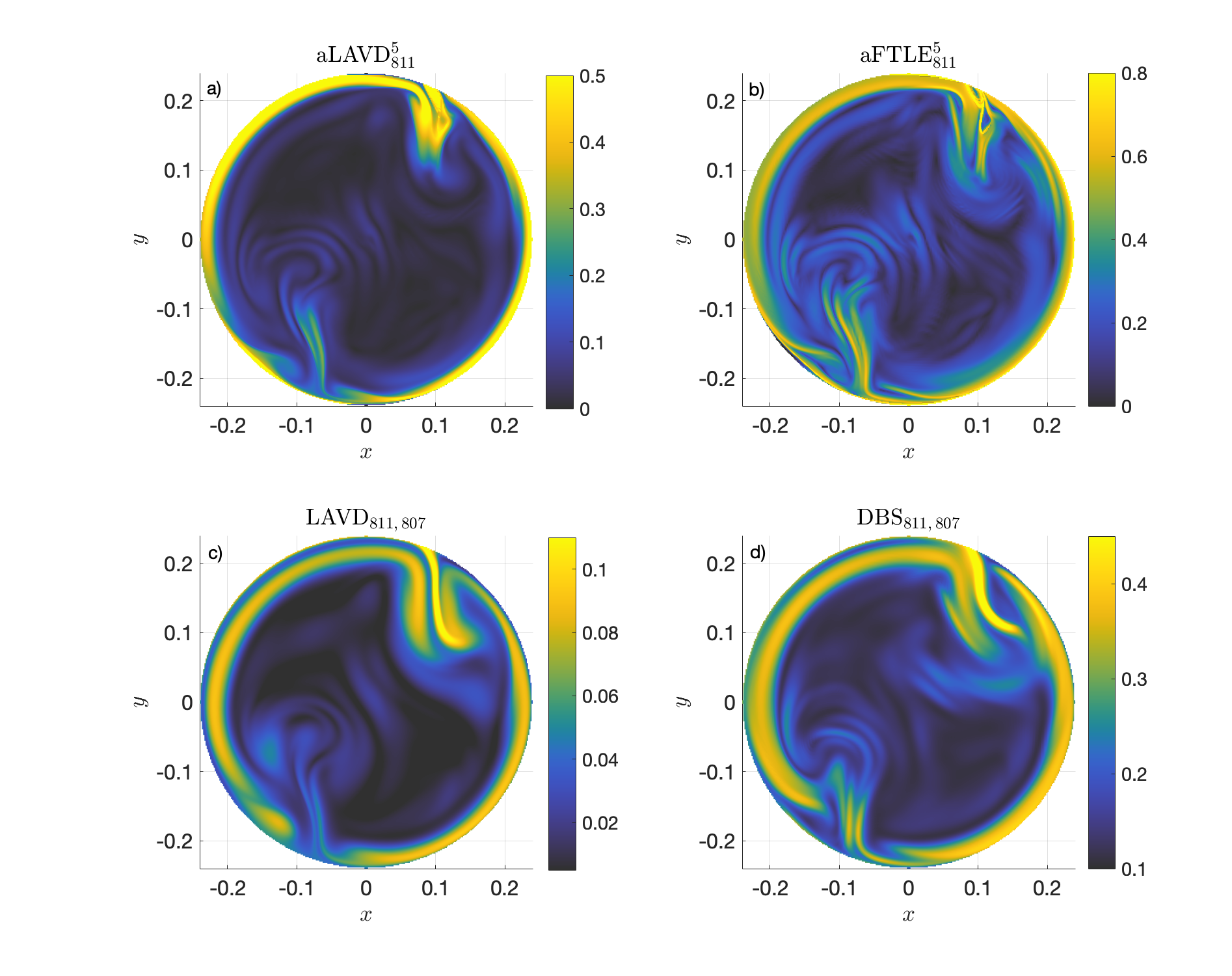}}% Images in 100% size
\caption{Comparison of advective, diffusive and active barriers for $\mathrm{Ro}^{-1} = 33$ at $t/t_{\!f\!\!f}=811$, highlighting the features of sidewall boundary flow due to convective wall modes.}
\label{fig:Ro 33}
\end{figure}

 \subsection{$\iRo=6.25$}

For the intermediate rotation case ($Ro^{-1}=6.25$), our transport barrier diagnostics still reveal a complex network of active, advective, and diffusive transport barriers. In figure \ref{fig:Ro 16}, we plot aLAVD, aFTLE, LAVD, and DBS fields at $z/H=0.1$ at the same dimensionless flow time ($t/t_{\!f\!\!f}=809$) as that visualized in figure \ref{fig:Theta_Fields}b, computed with the same advection timescales as in Section \ref{iRo=0}.

In \ref{fig:Ro 16}a and \ref{fig:Ro 16}b we see the three Ekman vortices act as instantaneous momentum barriers in aLAVD and aFTLE fields. Surrounding these prominent features are multiple momentum barriers that have a less visible effect on the spatial distribution of $\Theta$. Surrounding the core of the vortices, there are also momentum barriers that are limiting momentum transport near the cores, effectively separating core regions from each other. This behavior can be seen as convex ridges of aLAVD and aFTLE that parallel and contour the ridges that define the central cores. Along approximately $y=0.15$, we see one such momentum barriers that loosely agrees with the $\Theta=\left<\Theta\right>_\mathcal{H}$ contour, separating that zone by blocking momentum transport into the other Ekman vortices.

Figures \ref{fig:Ro 16}c and \ref{fig:Ro 16}d show the $x-y$ intersection with advective and diffusive transport barriers at this same height. These Lagrangian diagnostics show a clear dominance of the Ekman vortices, and much less of the small scale transport barrier behavior seen in the instantaneous momentum barrier diagnostics. One significant similarity here is the organization of LAVD and DBS contours that separate each Ekman vortex from each other and from the center of the cylinder. The extent of this separation, and the exact location of these barriers cannot be obtained from $\Theta$ alone in figure \ref{fig:Theta_Fields}b.

 \subsection{$\iRo=33$}

In our last example, we compare advective, diffusive, and active barriers for an RB flow with high rotational strength. In figure \ref{fig:Ro 33}, we plot aLAVD, aFTLE, LAVD, and DBS fields at $z/H=0.1$ at the same simulation time as that visualized in figure \ref{fig:Theta_Fields}c, computed with the same advection timescales as in the previous two sections.

In \ref{fig:Ro 33}a and \ref{fig:Ro 33}b momentum barriers with considerably less complexity than in the previous two simulations. These features largely parallel contours in \ref{fig:Theta_Fields}c. There are two major features that are related to sidewall boundary flow resulting from the high rate of rotation. Our two Lagrangian diagnostics reveal advective and diffusive barriers at the same locations with approximately the same level of complexity with some additional detail at the crossovers from the cold to the warm fraction of the wall mode. This is in direct contrast to the previous two simulations where there was a large-scale agreement between all barrier types, but at finer scales, many differences could be found. The only thing that changed between all simulations is the rotational strength, $\iRo$. It appears that the degree to which advective, diffusive, and active barriers all agree in rotating RB convection depends on the relative role of Coriolis and buoyancy forces. This could relate to the ratio of any mechanically stabilizing forces to the buoyancy which drives the convection.

\subsection{Bulk Agreement of Heat and Momentum Transport}
\label{sec: Bulk}
By varying the Rossby number, we are able to evaluate the role of purely-convective and Coriolis-influenced structures in rotating RB convection. In our non-rotating case, $\mathrm{Ro}^{-1}=0$, the flow consists of turbulent convective plumes. In the high rotational strength case, the Coriolis effect impedes the vertical mixing in the bulk and triggers the formation of a sidewall boundary flow.  In both cases, momentum and heat are transported following their respective transport barriers.

\begin{figure}
\centerline{\includegraphics[scale=.2]{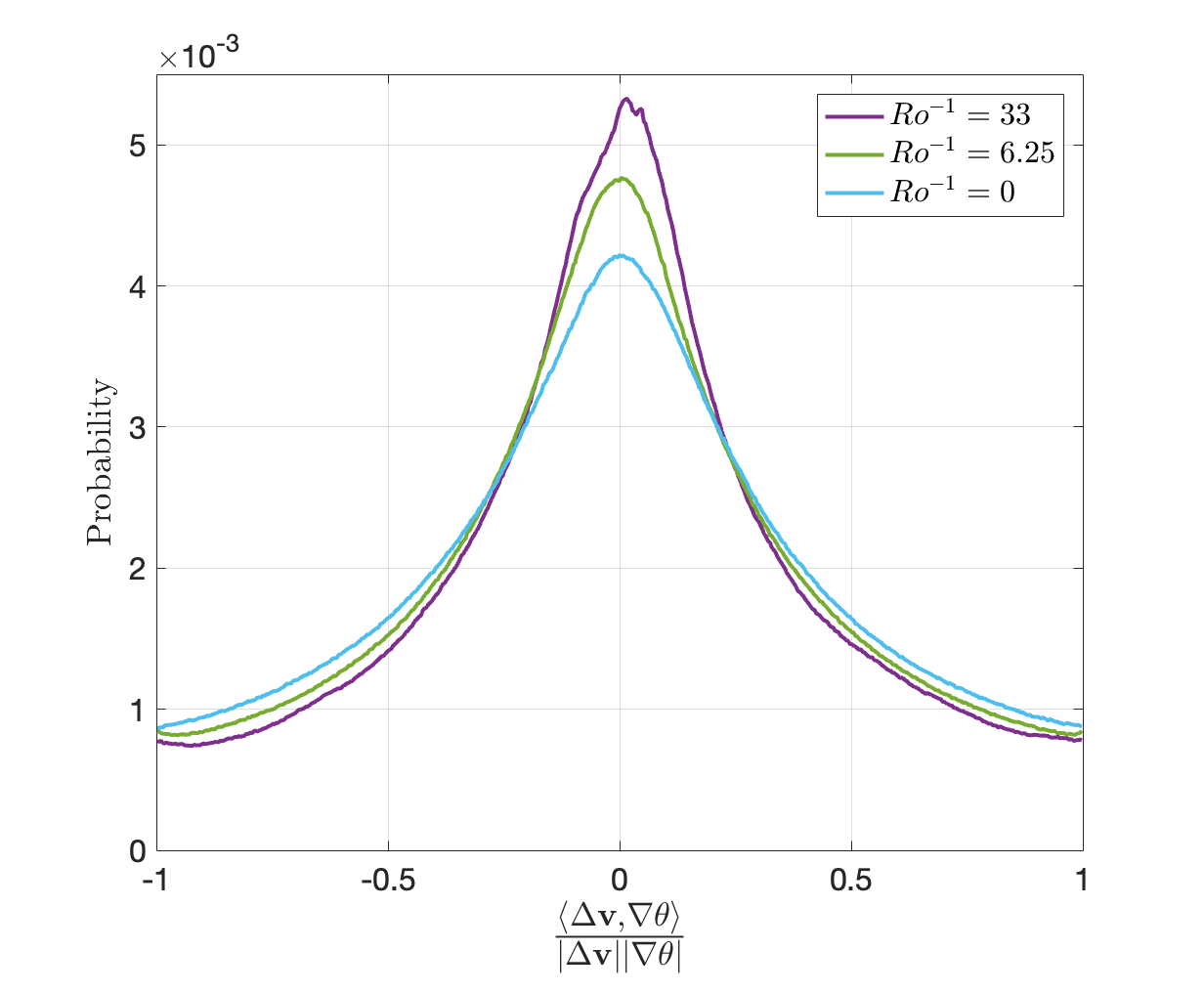}}% Images in 100% size
\caption{Momentum flux and heat flux inner products for the three strengths of rotation. Momentum and heat orthogonality increases as mechanical influence on flow structure dominates convective influence.}
\label{fig:Inner Products}
\end{figure}

The direction and magnitude of the diffusive transport of heat can be quantified by the gradient of $\Theta$. A transport barrier that effectively constrains the heat distribution in our flow will have a large temperature gradient across its boundary. These strong gradients are clearly visible surrounding the Ekman vortices in figure \ref{fig:Theta_Fields}c, which are responsible for the increase in the normalized heat transport in figure \ref{fig:NuVsRo}. Momentum transport barriers have been shown by \citet{Haller2020} to be stream surfaces in the barrier field (\ref{eq:active barrier}). Therefore, along a surface that acts as a transport barriers that constrains both the transport of momentum and heat, the inner product of the vectors $\langle \Delta \mathbf{v}, \nabla \theta \rangle$ is equal to zero. In figure \ref{fig:Inner Products}, we show the pointwise probability distribution of this inner product, after normalizing by vector lengths $|\Delta\mathbf{v}|$ \& $|\nabla \theta|$, for the entire flow volume over 26 free-fall time unit snapshots for each strength of rotation.

In all three flows we see a clear probability peak around zero, indicating a general agreement between momentum and heat transport barriers. As this calculation is applied for every grid cell, and not just along transport barrier surfaces, some variability around zero is to be expected. This probability indicates that the transport of momentum and heat is largely governed by similar features in the flow, though some disagreement exists. The alignment of heat and momentum transport increases as the Coriolis influence of rotation increases. We can see this as the variance in our bulk probabilities reduces for increasing $\iRo$, and the inner product distribution contracts around zero. In our complex RB flows, fluid motion is driven by both buoyant convection and rotational stabilizing forces. Their relative influence in turn changes the degree to which momentum and heat transport aligns, and how much their respective transport barriers agree. We investigate this balance further in Section \ref{Feedback}.

\section{Discussion}\label{discuss}

\subsection{Comparison of advection, heat, $\&$ momentum barriers}

 \begin{figure}
\centerline{\includegraphics[scale=.2]{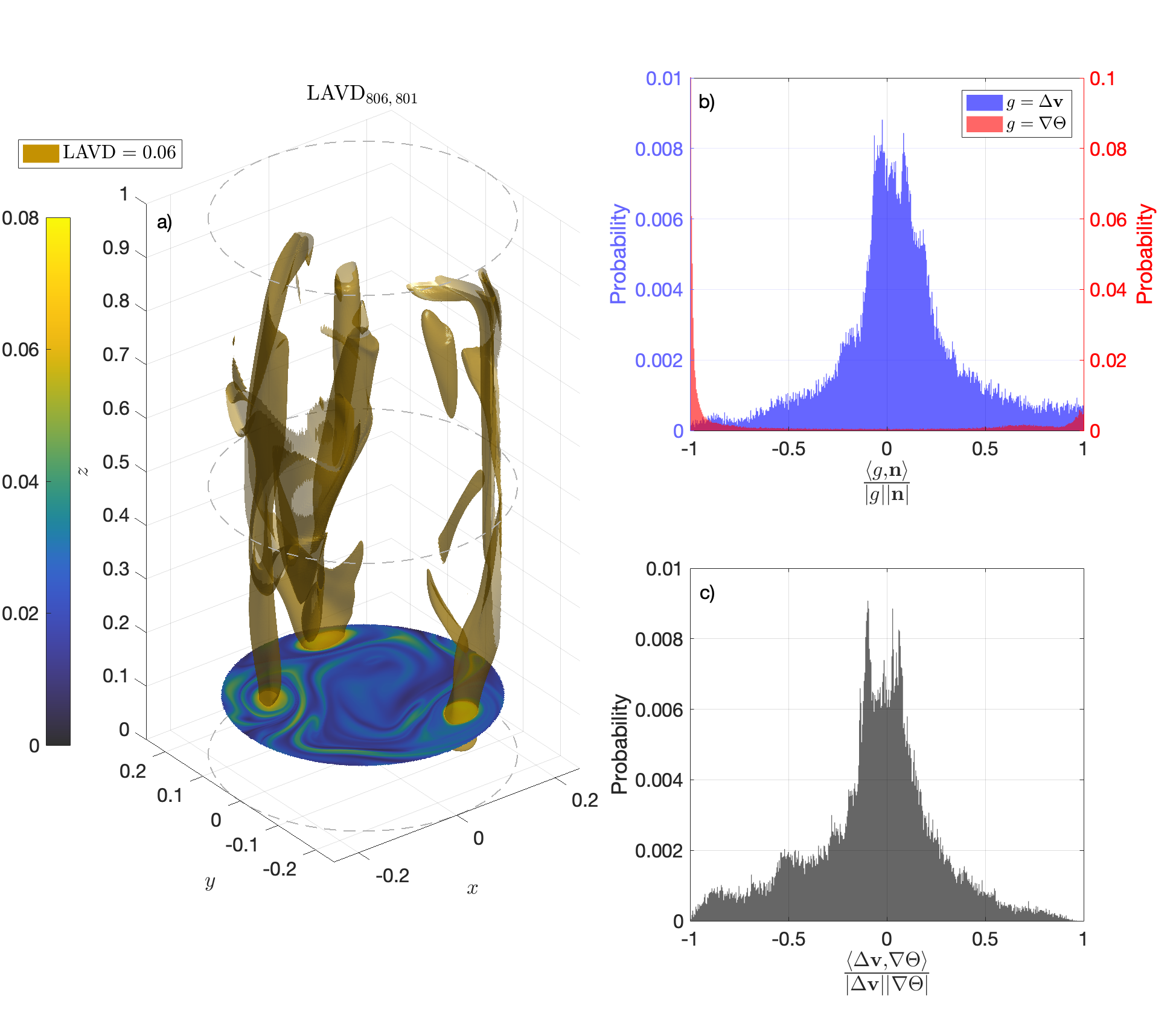}}% Images in 100% size
\caption{(a) Ekman vortices as LAVD isosurfaces (advective barriers). (b) Inner product of both diffusive heat flux and momentum flux through LAVD isosurfaces. (c) Inner product of diffusive heat flux and momentum flux vector along isosurface.}
\label{fig:Ekman Vortex}
\end{figure}

\begin{figure}
\centerline{\includegraphics[width=5in]{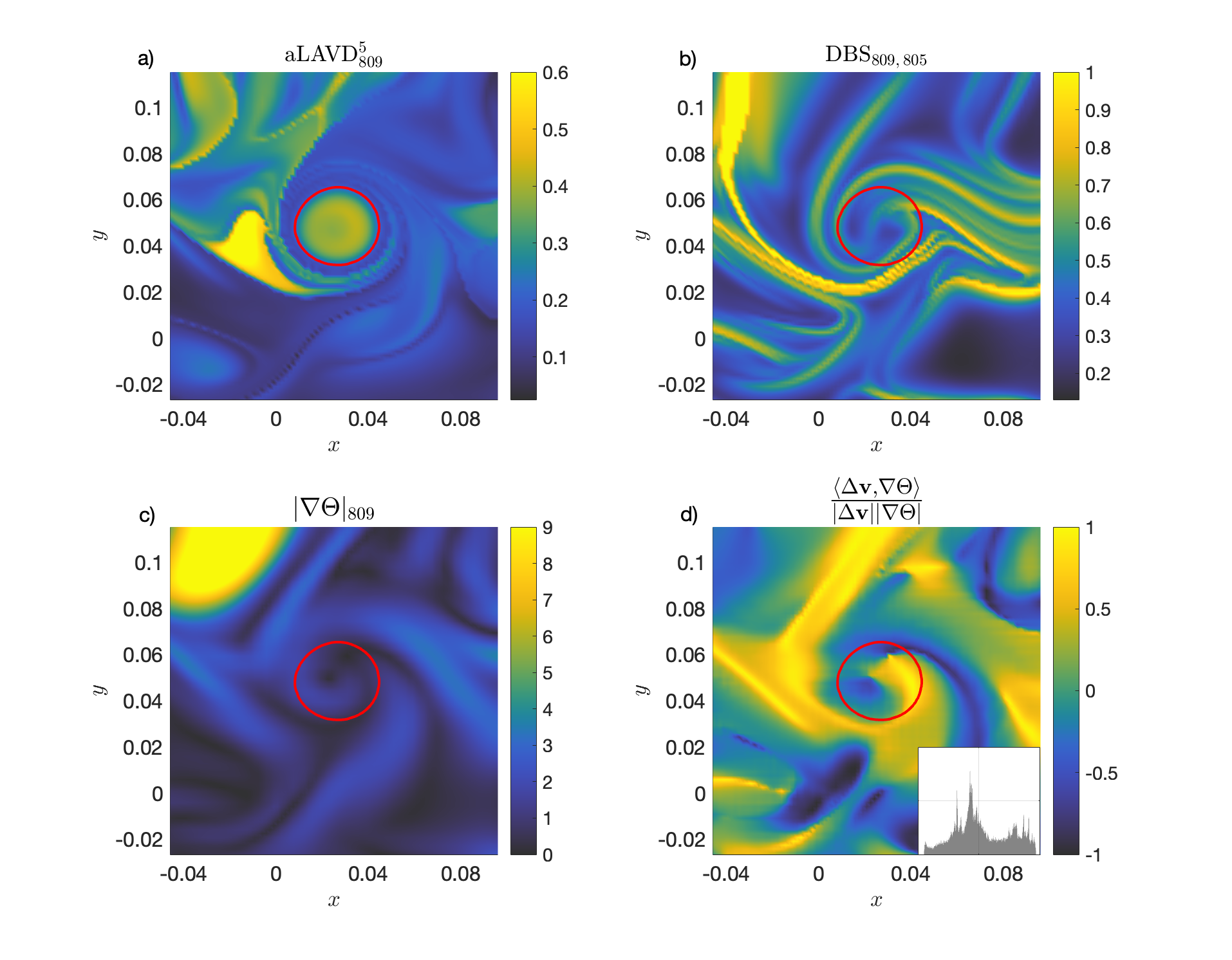}}% Images in 100% size
\caption{Comparison of active barriers and temperature fluxes. $\mathrm{Ro}^{-1} = 0$. Vortex leaking heat}
\label{fig:Leaky Vortex}
\end{figure}

 \begin{figure}
\centerline{\includegraphics[width=5in]{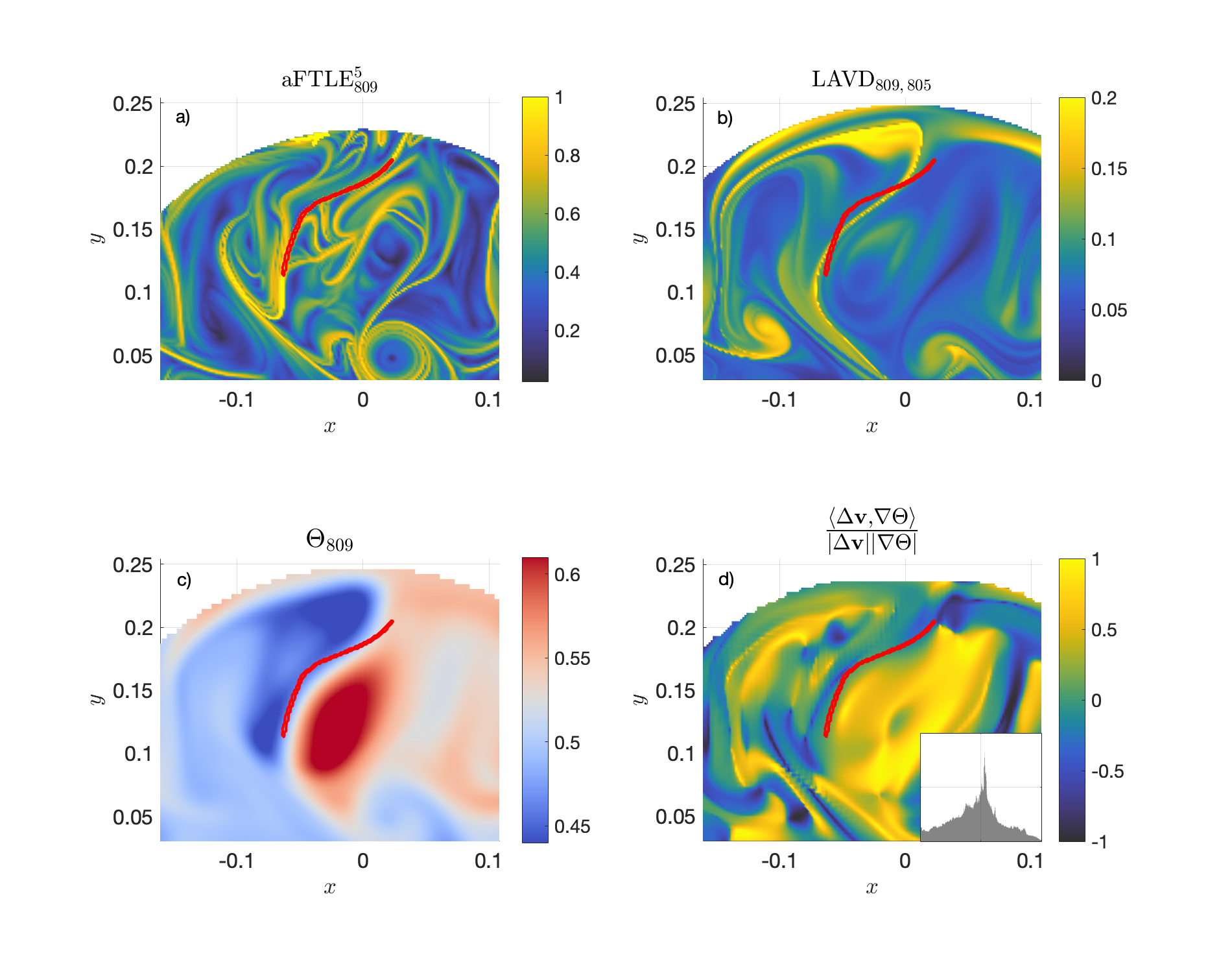}}% Images in 100% size
\caption{An example hyperbolic manifold in the momentum barrier field intersecting the $z=0.5$ plane represented as an aFTLE Ridge for the $\mathrm{Ro}^{-1} = 0$ flow. This momentum transport barrier is also behaving as a wall between cold and warm regions of the flow, thus maximizing diffusive heat transport.}
\label{fig:Blocking Wall}
\end{figure}

In the work of \citet{Haller2020}, DBS was mathematically derived such that DBS ridges highlight diffusive transport barriers to heat flux. In the original derivation, however, the authors were only able to verify this diffusive transport barrier behavior by looking at the evolution of an initial heat distribution that passively evolves with the flow. The present work is the first verification that DBS ridges indeed represent heat transport barriers that constrain the heat distributions and {\it modify} the flow.

To date, there has been limited investigation comparing the interplay of advective, diffusive, and active barriers, and no comparison for flows with both convective and mechanical forcing. It is exceptional to note that many of these barriers show ample agreement for our three cases, though they are all derived with distinct mathematical approaches. For example, in the intermediate rotation case with Ekman pumping ($\iRo=6.25$), the rotational Lagrangian coherent structures, seen as closed convex contours surrounding LAVD maxima, align precisely with the Ekman vortices in the temperature field, as well as the diffusion maximizing DBS ridges, and are parallel to the active momentum barriers in aLAVD and aFTLE. We make this comparison more rigorous by isolating one such LAVD level-surface and quantifying the flux of momentum and heat transport across it.

In figure \ref{fig:Ekman Vortex} we show how these three transport kinds of transport barriers align for the Ekman vortex case. Figure \ref{fig:Ekman Vortex}a shows the $\mathrm{LAVD}=0.06$ level surfaces in the entire flow volume, as it intersects the LAVD field at $z/H=0.1$. This approximation to more complex elliptic LCS extraction methods \citep[see][]{Neamtu-Halic2019} shows how robust LAVD fields are for identifying rotationally coherent three-dimensional transport barriers in time varying flows. In figure \ref{fig:Ekman Vortex}b, we calculate the orthogonality of momentum and heat flux vectors with surface-normal vectors on this LAVD isosurface. Even with our simplified LCS approach, we find near perfect orthogonality of diffusive heat flux with the rotationally coherent structure surface. The probability distribution obtains a peak value of 0.12 at -1 due to the hot-core of the Ekman vortices. Similarly, momentum flux is predominantly orthogonal to the surface-normals, revealing minimal leakage of momentum out of the Ekman vortices. 

Figure \ref{fig:Ekman Vortex}c compares the agreement of heat and momentum flux for every point on our advective barrier. We find on the surface of this elliptic LCS, there is also a strong pointwise agreement of heat and momentum transport. Though LAVD is a diagnostic field designed to identify elliptic LCS, and advective barriers, this example begins to rigorously show the interconnected nature of momentum, heat, and fluid advection barriers in RB flow. This also opens the door to evaluating the conditions under which momentum and heat fluxes are not in agreement.

As seen in figure \ref{fig:Inner Products}, the $\iRo=0$ case shows the greatest disagreement between momentum and heat flux vectors. We will now focus on examples of two structures with varying momentum and heat blocking properties. If we zoom in on the central region of the $z/H=0.5$ slice from Figs. \ref{fig:Theta_Fields}a and \ref{fig:Ro Inf}, we see what appears to be a cylindrical momentum barrier. We have isolated this region in figure \ref{fig:Leaky Vortex}. We extract the outermost closed convex contour surrounding the aLAVD maximum in the center of \ref{fig:Leaky Vortex}a. Though not highlighted, the aFTLE field reveals concentric closed ridges in the same region (Fig. \ref{fig:Ro Inf}b). 

When we compare this momentum blocking feature with the Lagrangian DBS field (Fig. \ref{fig:Leaky Vortex}b), instead of a closed curve, we see a plume wrapping in on itself, entraining surrounding fluid. The magnitude of the $\nabla\Theta$ in figure \ref{fig:Leaky Vortex}c shows that indeed there are relatively high gradient ridges parallel to the momentum barrier in red, but there are also level sets of constant temperature that are mixing in to the momentum core. Lastly, in figure \ref{fig:Leaky Vortex}d, we visualize the normalized momentum and heat flux inner product and compare to the location of the momentum vortex core. There is clear evidence of the swirling feature, but the direction of momentum transport and diffusive heat flux are actually parallel for one arm of the swirl, entering from the right. This is in direct contrast to the typical orthogonality of momentum and diffusive heat flux that we have seen for other barriers. 

Using our closed aLAVD contour as initial positions, we can also rigorously extract the stream surface by calculating trajectories in eq. (\ref{eq:active barrier}). This barrier stream surface perfectly blocks active momentum transport. On the surface, however, the temperature gradient follows a wide range of orientations, and is only orthogonal on part of the surface. Inset in figure \ref{fig:Leaky Vortex}d is the probability distribution of the heat-momentum inner products for this three-dimensional stream surface. We see many more spikes away from zero than in previous distributions, suggesting a more complex relationship the roles of heat and momentum for fluid organization in this region.

Though $\iRo=0$ shows the greatest relative disagreement between momentum and heat flux vectors for our three flows, there is still significant alignment between heat and momentum barriers in this turbulent flow. We highlight one such hyperbolic momentum barrier that acts as a wall for both momentum transport and temperature mixing. Shifting our focus to the dominant hot-cold interface in figure \ref{fig:Blocking Wall}, we identify an aFTLE ridge in red (figure \ref{fig:Blocking Wall}a) that shows close approximation to LAVD and $\Theta$ contours. While aFTLE and $\Theta$ are both Eulerian diagnostics, and LAVD is Lagrangian, we can see there is actually greater agreement between $\Theta$ (figure \ref{fig:Blocking Wall}c) and LAVD contours (figure \ref{fig:Blocking Wall}d) around our momentum barrier in red. Using this curve as initial conditions, we again calculate the full three-dimensional momentum barrier as a stream surface in eq. (\ref{eq:active barrier}). For this hyperbolic momentum barrier, we find a much clearer peak probability distribution of heat-momentum orthogonality in figure \ref{fig:Blocking Wall}d. This confirms our qualitative conclusion that our aFTLE ridge is representing an effective momentum and heat transport barrier.

\subsection{Feedback of convection generating momentum and organizing the flow}
\label{Feedback}

In Section \ref{sec: Bulk}, we found that the agreement between heat and momentum fluxes varies with the strength of rotation. Specifically, when fluid motion is dominated by mechanically-generated structures, the momentum and heat transport barriers are more closely aligned. We propose this is the result of a feedback mechanism of balancing forces that organize the flow.

When mechanical forces are strong, and convection plays a smaller role in generating motion, heat distributions are passively organized in the flow. In this way, momentum barriers that actively organize fluid motion are also responsible for the distribution of heat, and create features that effectively block scalar transport across them. This leads to momentum barriers with large temperature gradients across their boundaries. This effect can be see in figure \ref{fig:Inner Products} where higher rotational strength ($\mathrm{Ro}^{-1}=33$) generates stronger orthogonality of momentum and heat fluxes. 

As rotational strength in our flows decreases, the role of buoyancy increases. Thermal plumes in RB flows also generate momentum and thus influence the location and shape of momentum barriers. Fluid motion is subsequently constrained by the shape of momentum barriers, but the transport of momentum (and therefore its barriers) is not entirely governed by heat gradients. From our DBS calculations, and the theory surrounding it, we know  we can perfectly identify the diffusion maximizers solely from the velocity field. Thus, there is a feedback from the momentum barriers back into heat barrier formation by way of the velocity field. This heat-momentum feedback loop is most complex at low $\mathrm{Ro}^{-1}$ when momentum transport barriers are influencing the flow, but imperfectly defining advective and diffusive heat barriers.

\begin{figure}
\centerline{\includegraphics[width=6in]{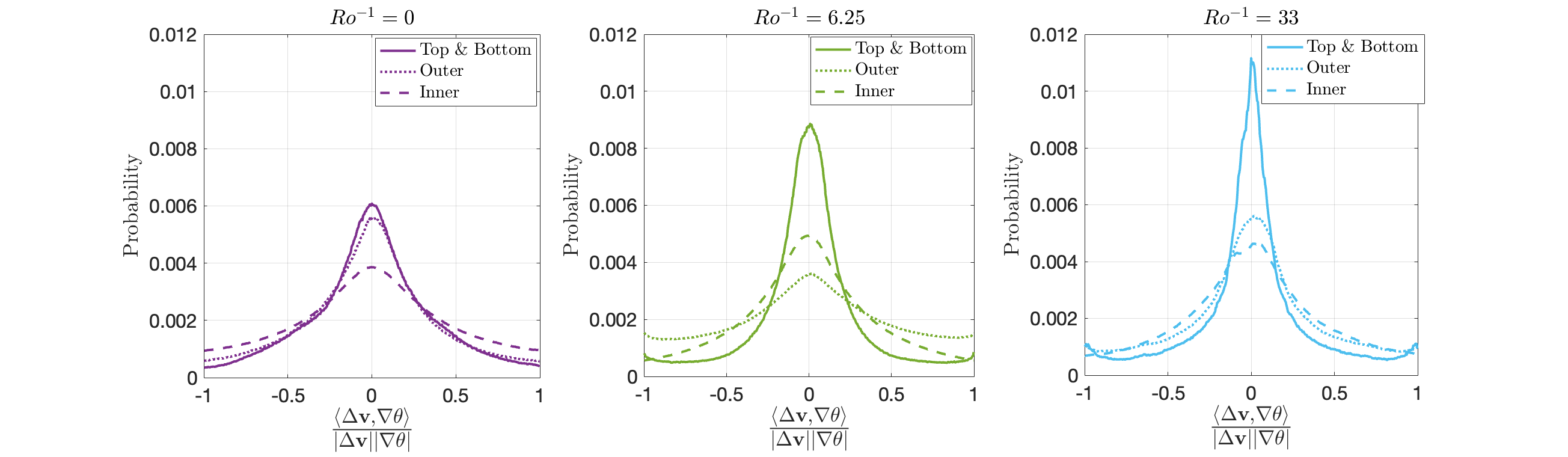}}% Images in 100% size
\caption{Momentum flux and heat flux inner products for the three strengths of rotation. Momentum and heat orthogonality increases as mechanical influence on flow structure dominates convective influence.}
\label{fig:Inner Products_Regional}
\end{figure}

As one final comparison, we consider separate fixed regions in our flow domain, and compare zones with varying influence from the walls. In figure \ref{fig:Inner Products_Regional}, we look at momentum and heat flux inner products near the top and bottom of the cylinder ($z/H<0.01 $ and $z/H>0.99$). For our three flows, we found the thermal boundary layer height $\lambda_T$ \citep{Stevens2010} is on the order of this range, equaling $1.11\times10^{-2}$, $9.37\times10^{-3}$ and $3.43\times10^{-2}$ for the three flows in order of increasing rotation strength. Sidewall boundary layer theory is less developed for weakly rotating RB flows, so we also looked at inner products in an outer region defined as the outermost $15\%$ of the volume ($r/H>0.23$), and took the complement of all these regions to be the inner region of the flow. We are limited as to how close we can get to the cylindrical walls by interpolated Cartesian grid used for the barrier diagnostics.

Momentum in ascending and descending fluid is redirected as that fluid encounters the upper and lower walls, respectively. As these hot and cold plumes deform along the shape of the walls of the flow domain, the deformation of diffusive barriers and momentum barriers is coupled. This results in a larger degree of agreement in active and diffusive barriers in the top and bottom regions for all three values of $\iRo$, where  orthogonality increases with increasing rotation strength. The peak probability for the top and bottom region inner products is larger than the respective bulk values in figure \ref{fig:Inner Products}. Though there is a strong thermal influence at the lower wall, inner product distributions are nearly identical if we further separate fluid near the top versus bottom (not pictured). This suggests we have isolated a zone with a particularly strong agreement between momentum and heat transport, resulting from a largely mechanical influence of the flow domain boundary.

In comparison, the inner region of the domain reveals significantly less agreement between momentum and heat transport, less than half the probability in the strong rotation case. The inner (and outer) domain, however, still shows a distinct peak probability of momentum and heat transport coupling. When we isolate outer wall and inner domains, trends in the probability distribution with increasing $\iRo$ are less clear.  This is especially evident for the outer wall distribution peak that actually decreases from $\iRo=0$ to $\iRo=6.25$ before increasing for $\iRo=33$. This requires further investigation, but may be due to the discretization of the round walls used for calculations of the velocity Laplacian, versus the original cylindrical coordinates used for the flow simulations. Using a different flow geometry, such as in a rectangular RB simulation would help answer this question and generate further comparison of the interplay between advective, diffusive, and active barriers. This may also be related to the specific role the Coriolis force is playing, enhancing thermal convection at $\iRo=6.25$ and suppressing it at $\iRo=33$.

\section{Conclusions and Outlook}\label{conclu}

We have shown that advective, diffusive, and active barriers reveal qualitatively similar structures in rotating RB flows for a wide range of rotational strength. Analysis of individual active and advective transport barrier surfaces quantifies a common agreement of flux-blocking behavior between various structures. Precise agreement of heat and momentum fluxes, and the exact shape of each respective transport barrier varies more with larger Rossby number. We propose this is connected to a heat-momentum feedback mechanism for competing forces that are influencing organization of the flow. The relative influence of mechanically and thermally-generated structures manifests as trends in the probability distributions of heat and momentum flux inner products for the entire flow domain. 

In addition to introducing these transport barrier identification methods to the study of Rayleigh-Bénard convection, this work has also utilized several novel tools for comparing the ability of structure to actually block momentum or heat transport. These metrics provide a new common ground for validating theories of zonal separation and tracking the individual structures governing the temporal evolution of heat flux.

Further investigation following this method of inquiry could improve our understanding of the underlying advective, diffusive, and active barriers that control the evolution of RB flows in far more complex geometries and scenarios. As a representative example of convective flows with analogs in many geophysical and industrial settings, the findings here suggest deeper insights into a variety of physical phenomena may be possible when analyzing coherent fluid structures with this dynamic transport barrier perspective. Many future applications exist, such as improving our understanding of the coupling of convective cells and cyclogensis in the atmosphere and the mixing of different molten materials in metallurgy or planetary cores.

\backsection[Funding]{N.A. and G.H  acknowledge financial support from Priority Program SPP 1881 (Turbulent Superstructures) of the German National Science Foundation (DFG). N.A. acknowledges financial support from the Swiss National Science Foundation (SNSF) Postdoc Mobility Fellowship Project P400P2 199190. R.H. and D.L. acknowledge funding by the ERC Starting Grant UltimateRB (No. 804283/ERC-2018-STG) and ERC Advanced Grant DDD (ERC-2016-ADG).}

\backsection[Acknowledgements]{We acknowledge PRACE for awarding us access to MareNostrum 4 based in Spain at the Barcelona Computing Center (BSC) under project 2020235589. This work was partially carried out on the Dutch national e-infrastructure with the support of SURF Cooperative. The authors would also like to thank R. Verzicco and R.J.A.M. Stevens for engaging discussions on this topic.}

\backsection[Declaration of interests]{The authors report no conflict of interest.}

\backsection[Author ORCIDs]{N. Aksamit, https://orcid.org/0000-0002-2610-7258; R. Hartmann, https://orcid.org/0000-0002-4860-0449; D. Lohse, https://orcid.org/0000-0003-4138-2255; G. Haller, https://orcid.org/0000-0003-1260-877X}

\backsection[Author contributions]{N.A., R.H., G.H., and D.L. were all involved in the conceptualization of the research, the analysis of the findings, and the writing of the manuscript. N.A. performed transport barrier computations and visualizations. R.H. performed the RB convection simulations and visualizations.}

\bibliographystyle{jfm}
\bibliography{references}
%Use of the above commands will create a bibliography using the .bib file. Shown below is a bibliography built from individual items.
\end{document}